\def\beg{\begin{equation}}
\def\eeq{\end{equation}}
\def\bega{\begin{eqnarray}}
\def\eeqa{\end{eqnarray}}
\def\sech{{\rm sech}}
\def\a{\alpha}

\def\r{\rho}

\def\G{\Gamma}

\def\ll{\left}
\def\rr{\right}
\documentclass[aps,pra,12pt]{revtex4}
\usepackage{graphicx}
\begin{document}
 \title{Quantum information entropies of the eigenstates and the coherent state of the P\"oschl-Teller potential}
 \author{Rajneesh Atre$^{1,2}$\footnote{atre@prl.ernet.in}, Anil Kumar$^{3}$, C. Nagaraja Kumar
 \footnote{cnkumar@pu.ac.in}$^{3}$\\
 and \\Prasanta K. Panigrahi \footnote{prasanta@prl.ernet.in} $^{1,2}$
 }
 \affiliation{$^1$ Physical Research Laboratory,
 Navrangpura, Ahmedabad-380 009, India\\
 $^2$ School of Physics, University of Hyderabad, Hyderabad-500 046, India\\
$^3$ Department of Physics, Panjab University, Chandigarh-160 014,
India}
 %\date{\today}
 %\maketitle

 \begin{abstract}
The position and momentum space information entropies, of the
ground state of the P\"oschl-Teller potential, are exactly
evaluated and are found to satisfy the bound, obtained by Beckner,
Bialynicki-Birula and Mycielski. These entropies for the first
excited state, for different strengths of the potential well, are
then numerically obtained. Interesting features of the entropy
densities, owing their origin to the excited  nature of the wave
functions, are graphically demonstrated. We then compute the
position space entropies of the coherent state of the
P\"oschl-Teller potential, which is known to show revival and
fractional revival. Time evolution of the coherent state reveals
many interesting patterns in the space-time flow of information
entropy.
 \end{abstract}
\pacs{xyz}
\maketitle
\vskip1cm
\newpage

\section{Introduction}

Information entropy plays a crucial role in a {\em{stronger}}
formulation of the uncertainty relations. The information
theoretic uncertainty relations were first conjectured by Everett
\cite{Everett} and Hirschman \cite{Hirschman} in 1957, and proved
by Bialynicki-Birula and Mycielski, and independently by Beckner
\cite{BBM}. From the general properties of the Fourier transform,
it was proved that, $S_{pos}+S_{mom} \ge 1+\ln\pi$.
%\bega \label{BBM} S_{pos}+S_{mom} \ge 1+\ln
%\pi \, . \eeqa
 Here, $S_{pos}=- \int{dx|\psi(x)|^{2}\ln|\psi(x)|^{2}}$ and $S_{mom}=-
\int{dp|{\tilde{\psi}(p)}|^{2}\ln|\tilde{\psi}(p)|^{2}}$, are the
position and momentum space entropies, respectively. In a $D$-
dimensional space, the right hand side of the above inequality
contains a multiplicative factor $D$. The above equation, known in
the literature as Beckner, Bialynicki-Birula and Mycielski (BBM)
inequality, captures the physical fact that localized
$|\psi(x)|^{2}$ leads to a diffused $|{\tilde{\psi}(p)}|^{2}$ and
vice-versa. It should be emphasized that, though $S_{pos}$ and
$S_{mom}$ are individually unbounded, their sum is bounded from
below. It is interesting to point out that, the above mentioned
inequality was discovered by Everett in the context of {\em{many
worlds interpretation}} of quantum mechanics. A general framework
for deriving uncertainty relations of the above type, between
general dynamical variables, not necessarily canonically conjugate
ones, have been given recently
\cite{Deutsch,Partovi,Birula1,Dodo,Uffink,Ruiz}.

These entropies lead to new and stronger version of the Heisenberg
uncertainty relations. Using a variational inequality, relating
entropy and standard deviation for an arbitrary one dimensional
variable $A$
 \cite{BBM,Shannon,Dehesa1}:
%\bega
 $S(A)\leq \frac{1}{2}+ \ln({\sqrt{2\pi} \Delta A})$ and the BBM inequality,
%\eeqa and Eq. (\ref{BBM}),
  one can derive Heisenberg type
uncertainty relations. These relations, which are based on
standard deviations, are not very reliable, particularly when
conjugate variables are discrete and the corresponding Hilbert
space is finite dimensional. These entropies have been quite
useful for characterizing quantum entanglement, since the von
Neumann entropy $ S=-Tr\{\hat{\r}\ln \hat{\r}\}, $ for the reduced
density operators, is often difficult to calculate. Various
properties of quantum mechanical entropy and its classical
counterparts have been elucidated in
Refs.\cite{Wehrl,Wehrl1,Petz}. As is clear, the single-particle
distribution densities, measuring the spread of the wave functions
in coordinate and momentum spaces, define their respective
entropies. Interestingly, in the density functional theory of
Hohenberg and Kohn, the single particle densities also completely
characterize a many-body system \cite{Kohn}.

The analytical determination of position and momentum space
entropies have been carried out only for a few quantum mechanical
systems. For the simple harmonic oscillator, the entropies were
exactly calculated for the ground state, in both, coordinate and
momentum space, for which the BBM inequality is saturated
\cite{opatrny}. For an arbitrary state, the entropies were
determined approximately, using asymptotic values of the entropy
of the orthogonal Hermite polynomials. The entropy integrals $\int
P^2_n(x)\ln P^2_n(x)d\mu(x)$, for several orthogonal polynomials
$P_n$'s having suitable measures, have been recently studied, from
which the asymptotic expressions for the information entropies for
large values of $n$, have been obtained for $D$ dimensional
harmonic oscillator and Coulomb problems \cite{Dehesa}.
Information entropy of neutral atoms \cite{Gadre,Gadre1}, in the
Thomas-Fermi theory, also manifests in a universal form, analogous
to the one given in Ref. \cite{BBM}. Information entropies in
various contexts e.g., mathematical physics, mathematics,
information theory, chemical physics and other areas of physics,
have been extensively analyzed in recent times
\cite{Majernik,Majernik1,Jacobi,Romano,Sheorey,Santha,Buyarov,Buyarov1,AbeSuzuki,Bala,GSA,Abe,
Barnett,Abe1,Coffey,Ferrari,Rajagopal}.

The present article is devoted to the study of the information
entropies of the P\"oschl-Teller (PT) family of potentials. These
potentials widely appear in the analysis of soliton bearing
nonlinear equations e.g., Bose-Einstein condensates and in quantum
problems on curved background \cite{Das,Pethick,Jatkar}. In the
following section, we first consider the hyperbolic PT potential
and evaluate the position and momentum space entropies exactly for
the ground state, for a range of potential strengths. For the
first excited state, we calculate these entropies, numerically,
which is shown to satisfy the bound obtained by Bialynicki-Birula
and Mycielski. Some interesting features of the entropy densities
are then graphically demonstrated. In the third section, we
compute the position space entropy densities of the coherent
states of the trigonometric PT potential, which exhibits revival
and fractional revival, due to interference effects. Under time
evolution these densities reveal interesting patterns in the
space-time flow of information entropy. We conclude in the last
section after pointing out various future directions of work in
this area.

\section{Information Entropy for P\"{o}schl-Teller systems }

 We begin with the Schr\"{o}dinger equation for hyperbolic
 P\"{o}schl-Teller potential, which is reflectionless and admits $n$ bound states for
integer values of $n$ (In units $\hbar=2m=1$), \bega \ll[ -
{d^2\over dx^2} - {n(n+1)\over 4} \sech^2({x\over2})\rr]
\psi^{(n)}(x) = E\psi^{(n)}(x) \quad . \eeqa

 The normalized ground state eigenfunction is given by,
\beg \label{normwavf}
\psi_o^{(n)}(x)={1\over\sqrt{2B({1\over2},n)}}\cdot \sech^n
({x\over2}) ~ , \eeq
 where $B({1\over2},n)$ is the Beta-function.

Using the definition of position space entropy, after a lengthy
but straightforward calculation, we obtain the analytical
expression: \beg
 S_{pos} = -(2n-1)\ln2+\ln
B\ll({1\over2},n\rr)+2n[\Psi(2n)-\Psi(n)]\,\,, \eeq
where $\Psi$ is the digamma function.\\
For $n=1$ and $2$, the position space entropy has the values
$S_{pos}=2$ and ${10\over3}-\ln 6$, respectively.

The corresponding momentum space entropies can be evaluated by
first obtaining the momentum space ground state wavefunctions,
which are the Fourier transforms of the corresponding position
space wavefunctions: \beg \psi_0^{(n)}(p)=A \, 2^{n} B
\ll({n\over2} + ip,{n\over2}-ip\rr) \, . \eeq
 For $n=1$, $\psi_o(p)=\sqrt{{\pi\over2}}$ $\sech (\pi p)$ and
the $S_{mom}$ can be easily evaluated,
\begin{eqnarray*}
S_{mom}&=& - \int_{-\infty}^{\infty} 2\psi_o^2(p)\ln\psi_o(p)dp\\
&=&2-\ln(2\pi)\, ,
\end{eqnarray*}
and the corresponding BBM inequality reads
$$
S_{pos} + S_{mom} = 4-\ln2\pi \geq 1+\ln(\pi).
$$
For higher values of $n$, evaluation of momentum space entropies
is quite cumbersome, instead we plot entropy-densities for both
position and momentum space. As shown in Fig.1 and 2, it is
interesting to notice that, the position entropy-density plots,
develop a dip at its peak as we increase the value of the
parameter $n$; exactly contrary behavior is observed in their
momentum space counterparts. As seen in Fig3., for the ground
state, as $n$ increases, the BBM inequality tends to be saturated.
Physically, for increasing $n$, the depth of the potential
increases and it increasingly resembles the oscillator potential,
which saturates the above inequality.

%----------------------------------------------------------------
%for one-column figure
\begin{figure*}
\begin{tabular}{ccc}
\scalebox{0.25}{\includegraphics{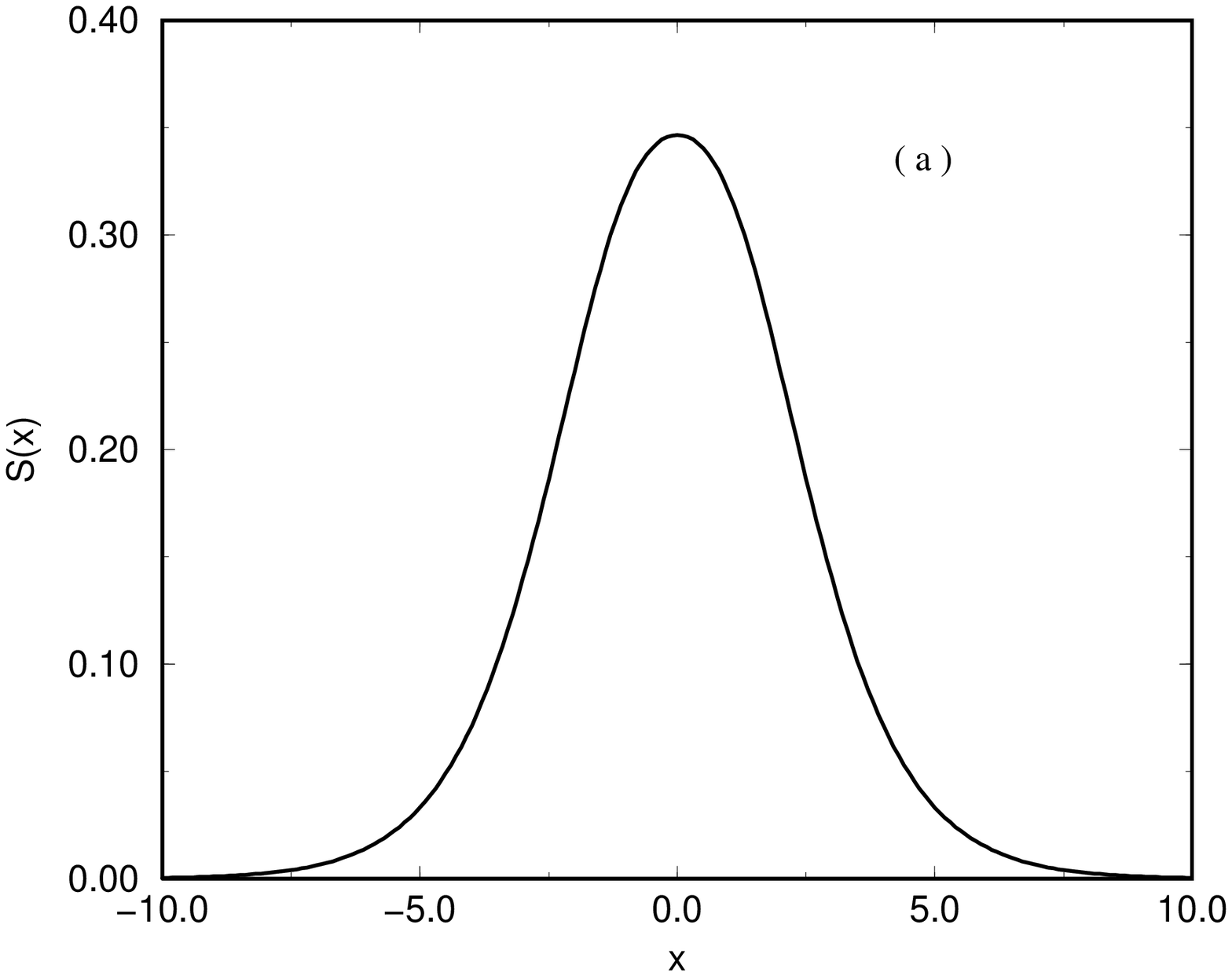}}&
\scalebox{0.25}{\includegraphics{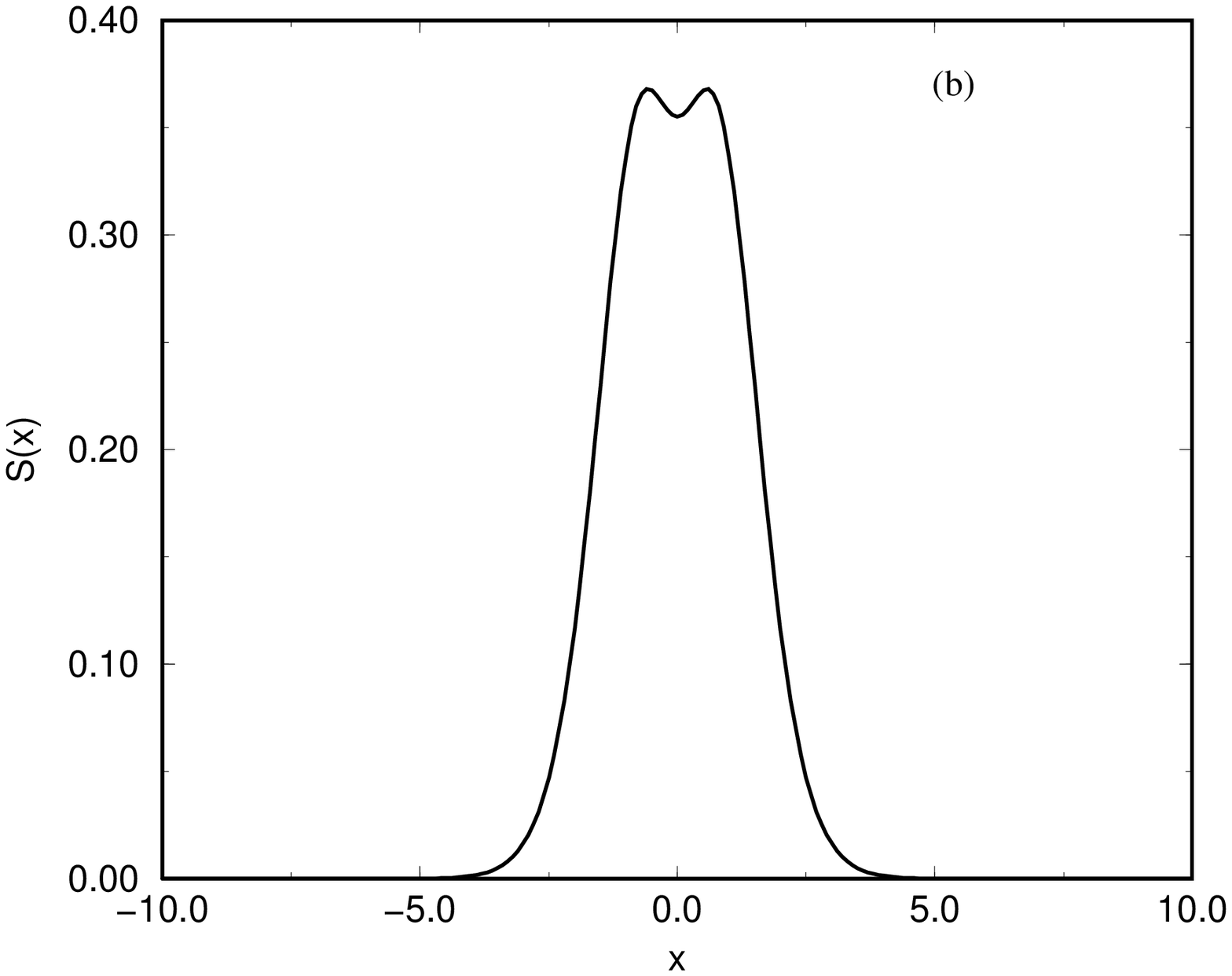}}&
\scalebox{0.25}{\includegraphics{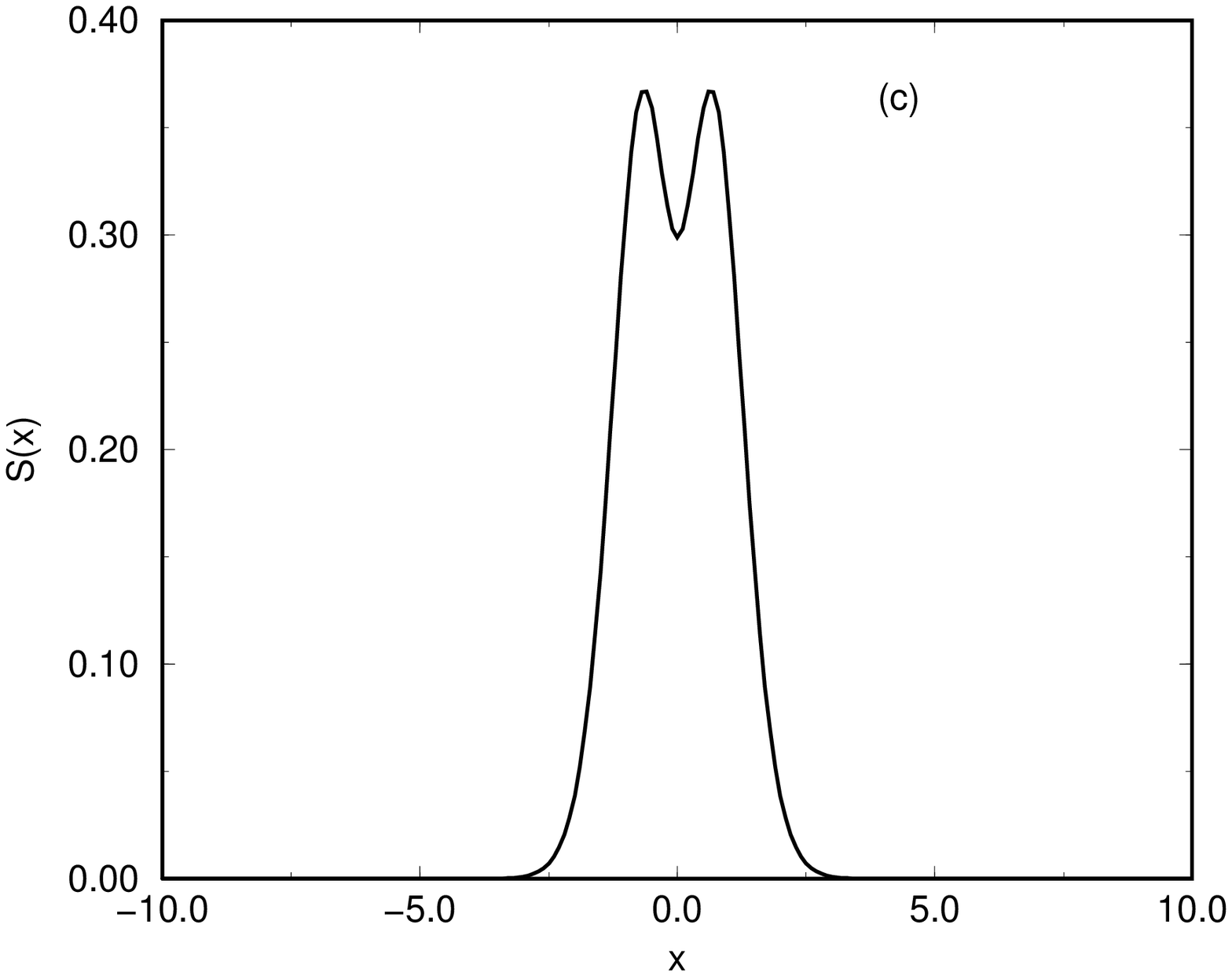}}
\end{tabular}
\caption{\label{SPT}Plots of the position space entropy densities
for the ground state of hyperbolic P\"oschl-Teller potential for
(a) $n=1$, (b) $n=3$ and (c) $n=5$.}
\end{figure*}
% for two-column figures
\begin{figure*}
\begin{tabular}{ccc}
\scalebox{0.25}{\includegraphics{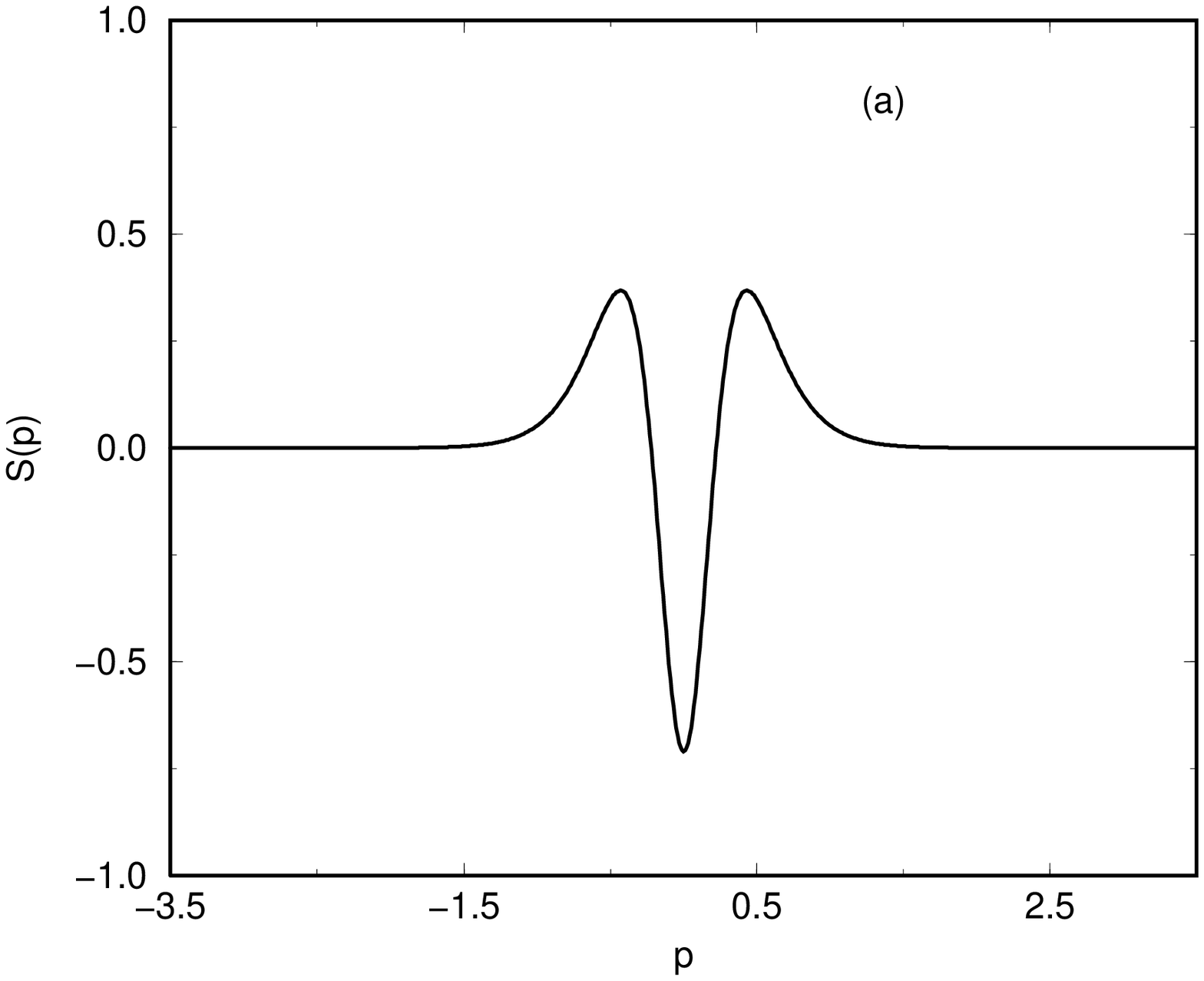}}&
\scalebox{0.25}{\includegraphics{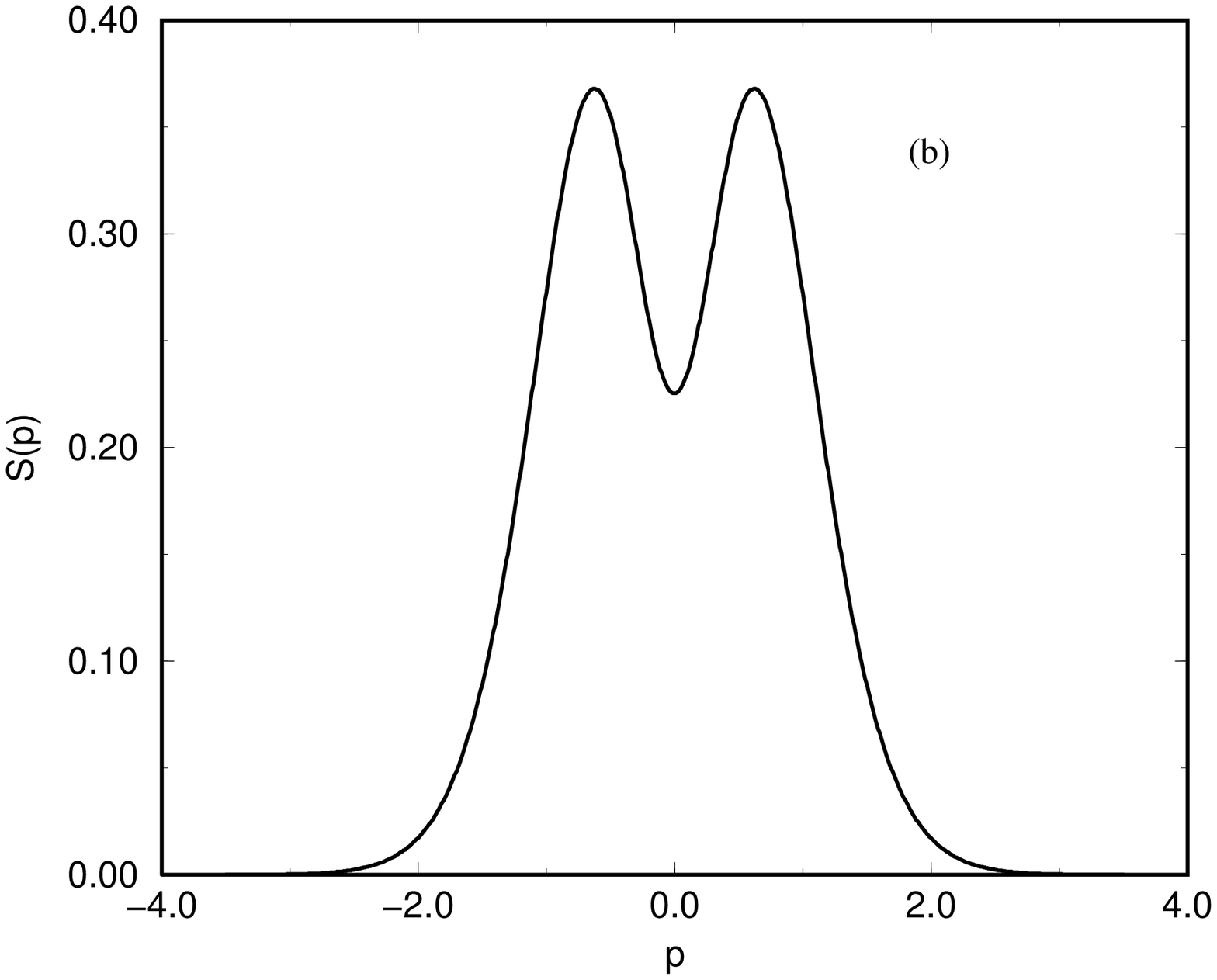}}&
\scalebox{0.25}{\includegraphics{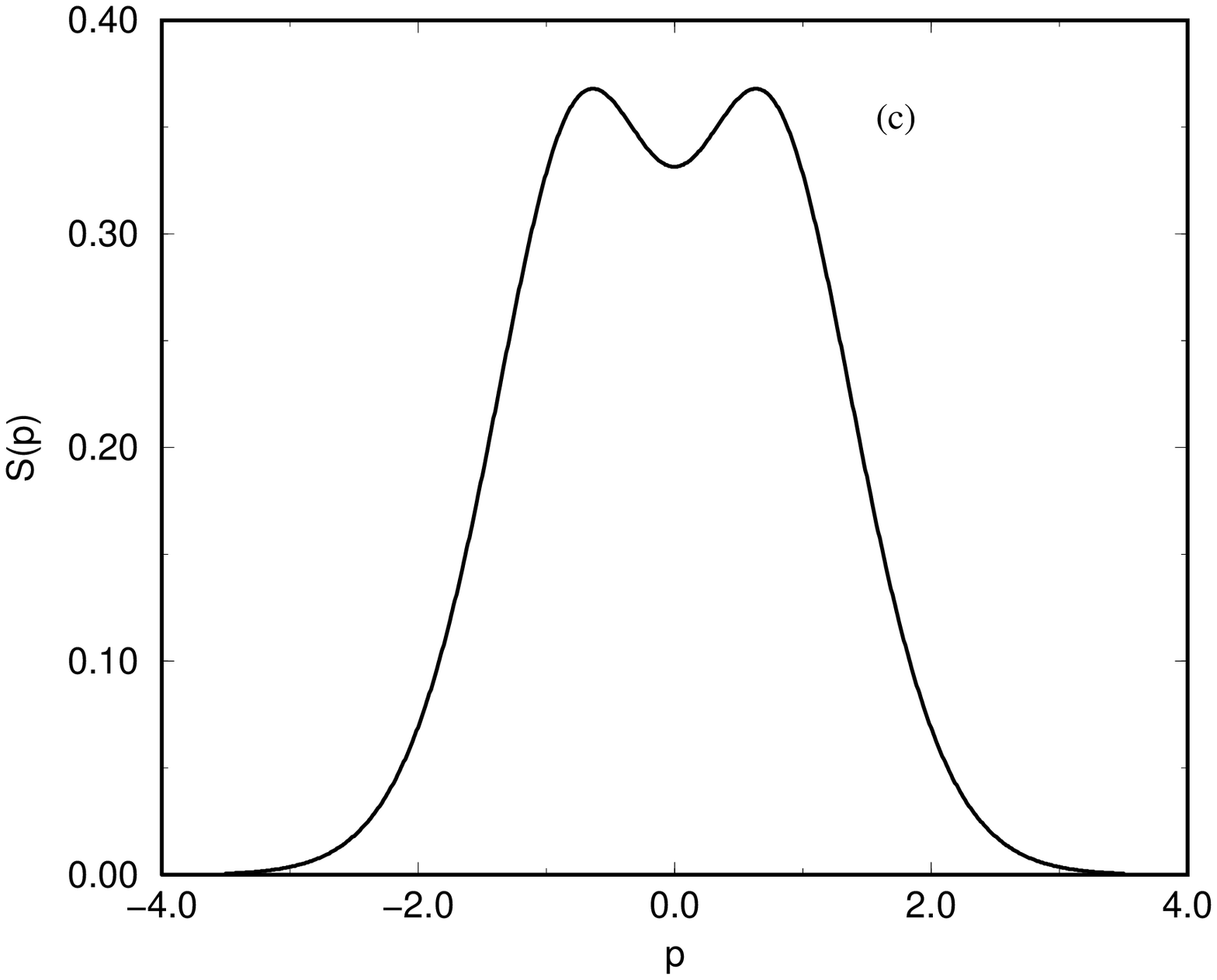}}
\end{tabular}
\caption{\label{SPT1}Plots of the momentum space entropy densities
for the ground state of hyperbolic P\"oschl-Teller potential for
(a) $n=1$, (b) $n=3$ and (c) $n=5$.}
\end{figure*}
%----------------------------------------------------------------
\begin{figure*}
%\begin{tabular}{c}
\scalebox{0.45}{\includegraphics{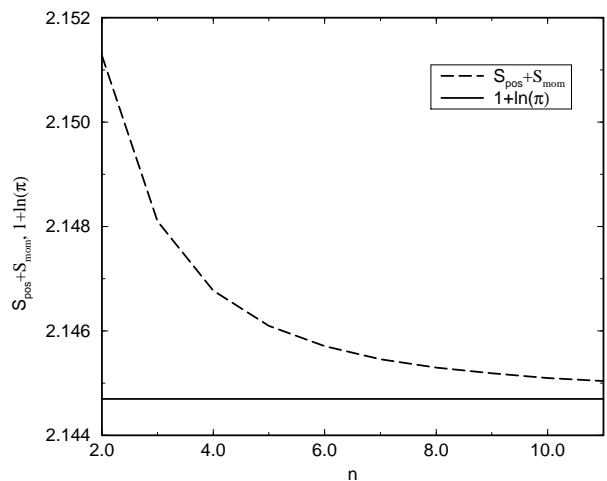}}
%\end{tabular}
\caption{\label{SPT}The Plot depicting the variation of sum of
position and momentum space entropies with respect to potential
parameter $n$.}
\end{figure*}
%----------------------------------------------------------------
We now proceed to evaluate the position space entropies of the
first excited state. The corresponding wave function for the
potential $V(x)$ reads,
\begin{equation}
\psi_1^{(n)}(x) = N\sech^{n-1}({x\over2})\tanh ({x\over2})
\qquad\mbox{for} \qquad n\ge2\,\,\,,
\end{equation}
where $N = {1\over\sqrt{2[B({1\over2},n-1)-B({1\over2},n)]}}$ is
the normalization constant.

For $n=2$, $S_{pos}=2.23472$ and for general $n$ the behavior of
the position space information entropy is depicted in Fig.2.
Table.1, depicts the BBM inequality for the first excited state as
a function of $n$. One sees that, as the value of $n$ increases
the sum of the entropies tends towards a saturation value higher
than the ground state value.
%--------------------------Table-----------------------
%\newpage
\begin{widetext}
\begin{center}
{\bf{Table 1: Table for BBM inequality for the first excited state
of the hyperbolic P\"oschl-Teller potential.}}
\end{center}

\begin{center}
\begin{tabular}{|c|c|c|c|c|c|c|c|c|c|}\hline
$n$&$S_{pos}$&$S_{mom}$&$S_{pos}+S_{mom}$&$1+\ln\pi$&$n$&$S_{pos}$&$S_{mom}$&$S_{pos}+S_{mom}$&$1+\ln\pi$\\
\hline &&&&&&&&&\\

$2$&$2.23472$&$0.722555$&$2.95728$&$2.1447$&$8$&$1.0971$&$1.63508$&$2.73217$&$2.1447$\\
&&&&&&&&&\\
$3$&$1.7988$&$1.0384$&$2.8372$&$2.1447$&$9$&$1.02621$&$1.70025$&$2.72646$&$2.1447$\\
&&&&&&&&&\\
$4$&$1.56242$&$1.22799$&$2.7904$&$2.1447$&10&$0.96409$&$1.7579$&$2.72199$&$2.1447$\\
&&&&&&&&&\\
$5$&$1.40082$&$1.36474$&$2.76556$&$2.1447$&11&$0.908807$&$1.80958$&$2.71839$&$2.1447$\\
&&&&&&&&&\\
$6$&$1.27825$&$1.47193$&$2.75018$&$2.1447$&12&$0.859009$&$1.85643$&$2.71544$&$2.1447$\\
&&&&&&&&&\\
$7$&$1.1796$&$1.56013$&$2.73973$&$2.1447$&13&$0.81371$&$1.89926$&$2.71297$&$2.1447$\\
&&&&&&&&&\\
\hline
\end{tabular}\\
\end{center}
\end{widetext}
%
%-----------------
\section{Entropy Densities for Coherent States of Trigonometric P\"oschl-Teller Potential}
Quantum systems with eigen spectra depending quadratically on a
quantum number $n$ are known to show revival and partial revivals
in time evolution of corresponding wave packets. These quantum
carpet structures have been studied quite extensively
\cite{Bluhm}. There have been suggestions to use the revival
structure for obtaining a factorization algorithm \cite{Mack}. The
possibility of realizing PT type of potentials in atomic systems
such as BEC, through optical means, makes the study of time
evolution of these systems more interesting \cite{Kutz}. It should
be pointed out that perturbation of BEC on a soliton or cnoidal
wave type solitary train background are known to satisfy the
hyperbolic or trigonometric PT Schr\"odinger equations
\cite{Pethick}. In the following we study the time evolution of
the information entropy density for an annihilation operator
coherent state of the trigonometric PT potential \cite{Charan}.
The fact that, coherent structure like laser is an annihilation
operator eigen state and the coherent manipulation of atoms,
possibly with optical means is being increasingly considered
seriously, may make these analyses useful. The trigonometric case
has been chosen deliberately, since it has an infinite number of
bound states as compared to the hyperbolic one, which makes the
construction of the coherent states straightforward.

We consider here the Hamiltonian of symmetric P\"oschl-Teller(SPT)
potential (in the units $\hbar=2m=1$), \beg
H=-\frac{d^2}{dy^2}+\frac{\alpha^{2}\r(\r-1)}{\cos^{2} (\a y)} ~,
 \eeq

 with eigenvalues and eigenfunctions, in the variable $x=\sin(\a y)$,

\bega E_{n}^{SPT}&=& \a^{2}(n+\r)^{2} \nonumber \\
\psi_{n}^{SPT}(x)&=&\left[\frac{\a(n!)(n+\r)\G(\r)\
\G(2\r)}{\sqrt{\pi}\G(\r+1/2)\G(n+2\r)}\right]^{1/2}(1-x^2)^{\r/2}C_{n}^{\r}(x)
\, .
\eeqa

 Recently based on a dynamical SU(1,1) algebra, an annihilation operator coherent state,
 was constructed for this system: $K_{-}|\gamma>=\gamma|\gamma>$, here $K_{-}$ is the annihilation operator
 of the SU(1,1) algebra \cite{Charan}. The coordinate space realization of this coherent state is given by:
\begin{equation} \label{CS}
\tilde{\chi}_{SPT}(x,\gamma)=N(\gamma)^{-1}
\sum_{n=0}^{\infty}{\ll[\frac{\Gamma(2\r)\Gamma(\r
+1/2){\sqrt{\pi}}}{\a(n!)(n+\r)\Gamma(2\r)\Gamma(2\r+n)}
\rr]^{1/2}\gamma^{n} \psi_{n}^{SPT}(x)}
\end{equation}

As already pointed out in the beginning of this section, the
quadratic nature of the spectra of SPT potential leads to the
possibility of revival and fractional revival in this quantum
system due to subtle interference effects. Keeping  in mind, the
fact that in the realistic situations the complete span of the
wave functions may not be available, we study the time evolution
of position space entropy densities for various values of $n$,
i.e., the number of states, interfering and constituting the
resultant coherent wave packet.  The effect of change of coherence
parameter $\gamma$ on the same is also analyzed. These are
depicted in the Figs. 4 and 5.  One finds dramatic changes in the
carpet structure of the entropy densities in space and time. One
observes rich tapestry like structures, where one can manipulate
the valleys and ridges of the entropy density in space and time.
It is interesting to observe that, as we increase the value of
coherence parameter $\gamma$, keeping $n$ fixed at some value, the
various ridges come close together and form a continuous
structure. These patterns become sharper for the higher values of
$n$.
%----------------------------------------------------------------
\begin{figure*}
\begin{tabular}{cc}
\scalebox{0.45}{\includegraphics{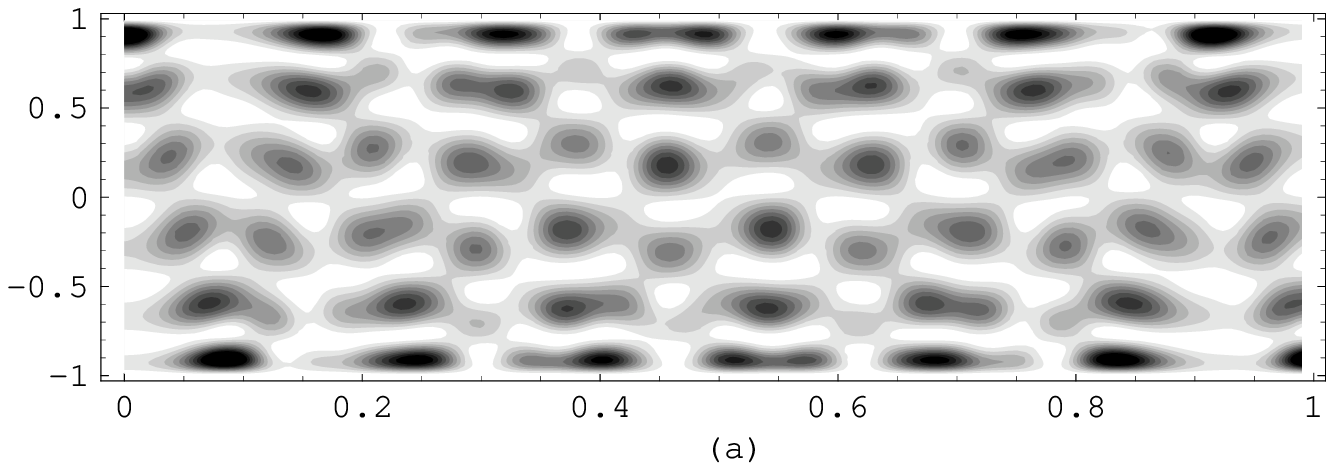}}&
\scalebox{0.45}{\includegraphics{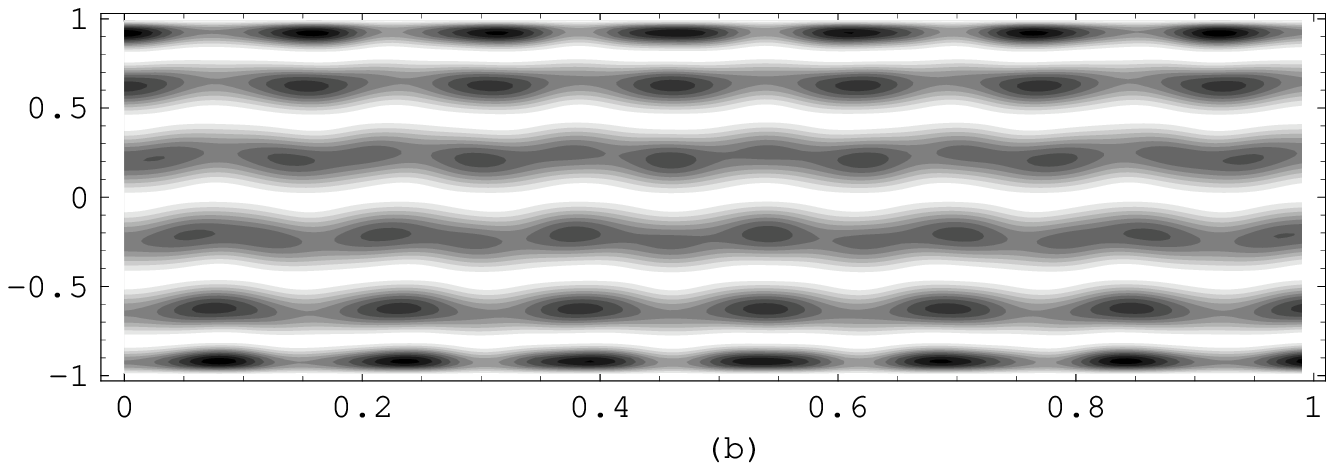}}
%\scalebox{0.35}{\includegraphics{Den5q30.eps}}
\end{tabular}
\caption{\label{SPT1}Contour plots, depicting time evolution of
position space entropy densities for the coherent states
trigonometric P\"oschl-Teller potential for (a) $n=5$, $\gamma=10$
and (b) $n=5$, $\gamma=30$. Darkness displays a low and brightness
a high functional value.}
\end{figure*}
%---------------------------------------------------------

\begin{figure*}
\begin{tabular}{cc}
\scalebox{0.45}{\includegraphics{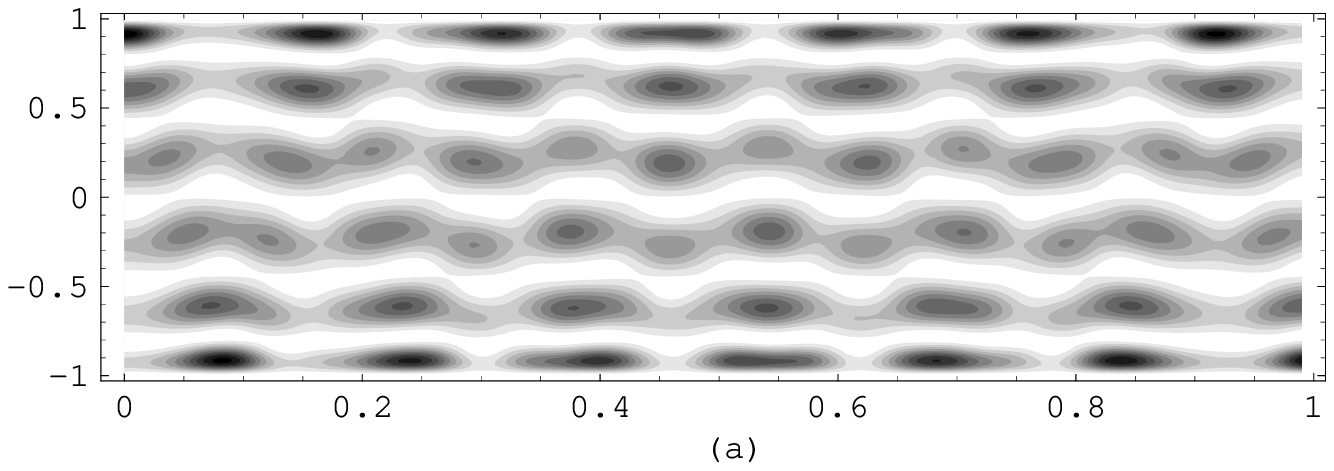}}&
\scalebox{0.45}{\includegraphics{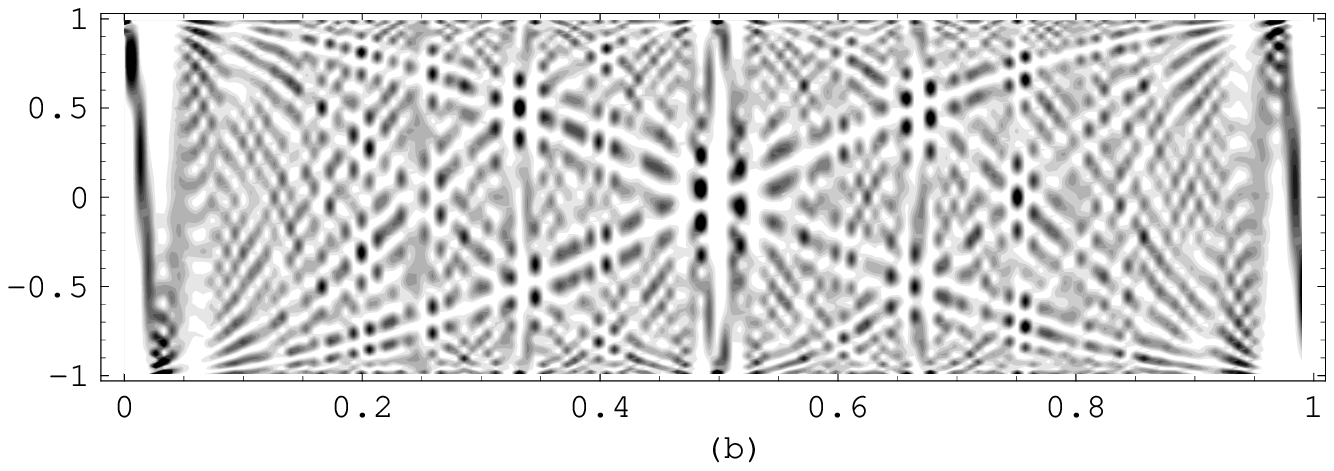}}
%\scalebox{0.35}{\includegraphics{Den5q30.eps}}
\end{tabular}
\caption{\label{SPT1}Contour plots, depicting time evolution of
position space entropy densities for the coherent states
trigonometric P\"oschl-Teller potential for (a) $n=5$, $\gamma=15$
and (b) $n=30$, $\gamma=15$. Darkness displays a low and
brightness a high functional value.}
\end{figure*}

%----------------------------------------------------------------

\section{Conclusions}
  In conclusion, we have studied the information entropies of a class of quantum systems
  belonging to the P\"oschl-Teller family of potentials. Exact results,
  for the position space entropies of the ground and first excited states of the hyperbolic
  P\"oschl-Teller potential were obtained, for a range of potential strengths. The expression for momentum
  space entropy was obtained analytically for the ground state and numerically
  computed for the first excited state.
It was found that, these entropies satisfy  the Beckner,
Bialynicki-Birula and  Mycielski inequality. The entropy densities
for the above cases were depicted graphically, for demonstrating
the entropy distribution in the well. For the trigonometric case,
after investigating the entropies associated with the eigenstates,
we studied the time evolution of entropy density  for the coherent
state \cite{Charan}. The intricate carpet structure shows the
richness of this quantum system, which needs to be explored
further. It should be noted that, coherent states are being
envisaged for the storage of quantum information. P\"oschl-Teller
potential manifests in quantum problems on curved background
\cite{Jatkar}, as also in non-linear integrable models with
soliton solutions like, KdV equation \cite{Das}. In light of this,
the physical relevance of the information entropies computed here,
needs further study. We hope to come back to some of these
questions in future.

{\bf{Acknowledgements:}} We acknowledge many useful discussions
with Prof. K.D. Sen, who also brought to our notice many relevant
references. We are thankful to Prof. V.B. Sheorey for discussions
and to Dr. J. Banerji for help in the numerical works. A.K.
acknowledges the financial support of CSIR India through S.R.F.

%\end{enumerate}

\end{document}